\documentclass[conference]{IEEEtran}
\usepackage{amsfonts}
\usepackage{amssymb}
\usepackage{cite}
\usepackage[cmex10]{amsmath}
\usepackage{float}
\usepackage{color}
\usepackage{stfloats,fancyhdr}
\usepackage{amsmath}
\usepackage{algorithm}
\usepackage{algorithmic}
\usepackage{multirow}
\usepackage{changepage}
\usepackage[normalem]{ulem}
\usepackage{amsmath}

\IEEEoverridecommandlockouts

\ifCLASSINFOpdf
  \usepackage[pdftex]{graphicx}
  \DeclareGraphicsExtensions{.pdf,.jpeg,.png}
\else
  \usepackage[dvips]{graphicx}
  \DeclareGraphicsExtensions{.pdf}
\fi

\usepackage{subfigure}
\usepackage{graphicx}
\usepackage{xcolor}
\usepackage{fancybox,dashbox}
\usepackage{float}

\begin{document}
%
\title{{On the Age of Information of Short-Packet Communications with Packet Management}
}
\author{\IEEEauthorblockN{Rui Wang\IEEEauthorrefmark{1}, Yifan Gu\IEEEauthorrefmark{1}, He Chen\IEEEauthorrefmark{2}, Yonghui Li\IEEEauthorrefmark{1}, Branka Vucetic\IEEEauthorrefmark{1}}%
	\IEEEauthorblockA{\IEEEauthorrefmark{1}School of Electrical and Information Engineering, The University of Sydney, Sydney, Australia}
	\IEEEauthorblockA{\IEEEauthorrefmark{2}Department of Information Engineering, The Chinese University of Hong Kong, Hong Kong SAR, China}
	\IEEEauthorblockA{
		\IEEEauthorrefmark{1}rwan7881@uni.sydney.edu.au, \{yifan.gu, yonghui.li, branka.vucetic\}@sydney.edu.au,
		\IEEEauthorrefmark{2}he.chen@ie.cuhk.edu.hk}
}

\maketitle

\begin{abstract}
In this paper, we consider a point-to-point wireless communication system. The source monitors a physical process and generates status update packets according to a Poisson process. The packets are transmitted to the destination by using finite blocklength coding to update the status (e.g., temperature, speed, position) of the monitored process. In some applications, such as real-time monitoring and tracking, the timeliness of the status updates is critical since the users are interested in the latest condition of the process. The timeliness of the status updates can be reflected by a recently proposed metric, termed the age of information (AoI). We focus on the packet management policies for the considered system. Specifically, the preemption and discard of status updates are important to decrease the AoI. For example, it is meaningless to transmit stale status updates when a new status update is generated. We propose three packet management schemes in the transmission between the source and the destination, namely non-preemption (NP), preemption (PR) and retransmission (RT) schemes. We derive closed-form expressions of the average AoI for the proposed three schemes. Based on the derived analytical expressions of the average AoI, we further minimize the average AoI by optimizing the packet blocklength for each status update. Simulation results are provided to validate our theoretical analysis, which further show that the proposed schemes can outperform each other for different system setups, and the proposed schemes considerably outperform the existing ones without packet management at medium to high generation rate.
\end{abstract}


\IEEEpeerreviewmaketitle

\section{Introduction}
Many emerging information systems require real-time monitoring and tracking. For example, in factory automation systems, the sensor generates status updates which measure the status (e.g., temperature, speed and position) of the machine, and transmits the status updates to a destination for control and tracking purposes. In such scenarios, the timeliness of the information is critical because the destination is only interested in the latest condition of the machine. In fact, the conventional performance metrics, such as throughput and delay, cannot properly characterize the timeliness of the status updates. The latter can be reflected by a recently proposed performance metric, named the age of information (AoI) \cite{AoI_FCFS}. The AoI captures both the latency and the generation time of each status update. Specifically, AoI is defined as the time elapsed since the generation of the last successfully received status update. If the most recently received status update carries the data sampled at time $r\left(t\right)$, the age of status update at time $t$ is defined as $t-r\left(t\right)$ \cite{AoI_FCFS}.

AoI has attracted a large amount of research interests very recently. The authors in \cite{AoI_FCFS} first studied the average AoI for a First-Come-First-Serve (FCFS) system by considering three different queueing models, named $M/M/1$, $M/D/1$ and $D/M/1$, respectively. The authors in \cite{AoI_LCFS} studied the Last-Come-First-Serve (LCFS) system and showed that the LCFS model achieves lower average AoI compared with the FCFS model by transmitting the latest status update first. The average AoI of more complex queueing models, $M/G/1$ and $M/G/1/2$ were considered in \cite{LHuang_MG1} and \cite{AEphremides_MM12_deadline}, respectively. In order to reduce the AoI and increase the reliability, different retransmission schemes were studied in \cite{AOIerror,YIFANAOI,MG11HARQ,AoI_HARQ}. Specifically, \cite{AOIerror,YIFANAOI} considered a classical automatic repeat request (ARQ) scheme, where the same information is retransmitted when it is received with error. Differently, \cite{MG11HARQ} and \cite{AoI_HARQ} studied the hybrid ARQ (HARQ) schemes. To improve the spectrum efficiency, there have also been a few investigations into the AoI of cognitive radio networks \cite{AValehi_CR,SLeng_CR,YifanAoICR}. Finally, the AoI of energy harvesting devices were analyzed in  \cite{AoI_EH1,AoI_EH2}.

The AoI metric under short packet communication were studied in \cite{AoISP,AoISP1,AoISP2}. This is motivated by the fact that the systems that can benefit from the AoI metric are usually monitoring and automation systems, with status updates of limited sizes. In conventional communication systems using the Shannon theory, errors depend on the received signal-to-noise ratio (SNR) and the coding rate. More importantly, an arbitrary small error probability can be achieved if the coding rate is below the Shannon capacity. Differently, in the short packet communication, errors always exist even the coding rate is below the Shannon capacity, and the error rate depends heavily on the packet blocklength. A natural question arises: what is the impact of packet blocklength on the AoI performance of a short packet communication system with status updates? \cite{AoISP,AoISP1,AoISP2} answered this question and concluded that there exists an optimal packet blocklength that can minimize the AoI metrics.
However, we realize that the packet management strategies, i.e., packet preemption and discard, was not considered in \cite{AoISP,AoISP1,AoISP2}. Specifically, the preemption of new status updates and the discard of stale status updates can reduce the AoI significantly because the destination is only interested in the latest status. In this paper, different from \cite{AoISP,AoISP1,AoISP2} that considered an FCFS queue with an infinite-sized buffer, we assume that the buffer can only store one status update and focus on the analysis of various packet management policies. Note that the considered buffer model is relevant to status update systems. It can maintain low cost, and it is meaningless to store outdated status updates when a new status update is generated. In the considered system, the new generated status update can either be discarded by the source or preempt the service of the current status update stored in the buffer. Besides, retransmission can also be introduced to the considered system to further decrease the AoI. We thus propose three packet management schemes, namely non-preemption (NP), preemption (PR) and retransmission (RT).

The main contributions of this paper are summarized as follows: (\text{1}) By considering a point-to-point system with a buffer that can store one status update, we propose three different packet management schemes, NP, PR and RT, to deliver the generated status updates as timely as possible. (\text{2}) We derive closed-form expressions of the average AoI for the proposed schemes under the finite blocklength regime. Based on the derived expressions, we further minimize the average AoI by optimizing the packet blocklength of each status update. (\text{3}) Simulation results are then provided to validate the theoretical analysis, which show that the proposed schemes can outperform each other. More importantly, they considerably outperform the existing ones using FCFS without package management when the generation rate of source is high.

\section{System Model}
\subsection{System Description}
We consider a point-to-point wireless communication network consists of one source and one destination. It is assumed that each generated status update at the source contains fixed $N$ bits of information, e.g., temperature, speed and position of the monitored physical process. The status updates contain packets with limited sizes and we use the terms ``status updates" and ``packets" interchangeably hereafter. The status update is coded into a signal with $m$ channel uses (c.u.), i.e., symbols. As in \cite{AoISP}, we assume an additive white Gaussian noise (AWGN) channel between the source and the destination. According to \cite{FB}, the block error rate (BLER) for AWGN channel using finite blocklength coding can be approximated as
\begin{equation}\label{BLER}
\varepsilon  \approx Q\left( {{{{1 \over 2}{{\log }_2}\left( {1 + \gamma } \right) - {N \over m}} \over {{{\log }_2}\left( e \right)\sqrt {{1 \over {2m}}\left( {1 - {1 \over {{{\left( {1 + \gamma } \right)}^2}}}} \right)} }}} \right),
\end{equation}
where the expression is very tight for $m>100$. $\gamma$ is the received signal-to-noise (SNR) ratio at the destination and $Q\left( x \right) = \int_x^\infty  {{1 \over {\sqrt {2\pi } }}{e^{ - {{{u^2}} \over 2}}}du} $ is the $Q$-function. It is worth mentioning that the analysis of average AoI provided in this paper can be readily extended to the fading channel case by replacing the expression of BLER in (\ref{BLER}) with that of a fading channel derived in \cite{yifanFB1, yifanFB2}. For the purpose of exploration, we consider an AWGN channel in this paper.
\subsection{Packet Management Schemes}
As in \cite{AoISP,AoISP1}, we assume that the status updates are generated at the source according to a Poisson process with rate $\lambda$. We assume that the information buffer at the source can only store one status update. Besides, to maintain low cost and implementation simplicity, we assume that there is no feedback signal provided by the destination. We subsequently investigate three different packet management schemes, namely NP, PR and RT. We now describe the operating principles of the three schemes in the following.
\subsubsection{Non-preemption (NP) scheme}
In the NP scheme, when a new status update is generated and the buffer is empty, the source starts the service of the generated status update. Otherwise, if a new status update is generated at the source and it is currently serving a status update, the source discards the new status update and keeps serving the current one. Besides, the source only sends the status update once and there is no retransmission.
\subsubsection{Preemption (PR) scheme}
Preemption is considered in the PR scheme such that new generated status update always preempts the service of the current status update. Specifically, if a new status update is generated at the source and the source is currently serving a status update, it discards the current status update and starts to serve the new one. There is also no retransmission in the PR scheme.
\subsubsection{Retransmission (RT) scheme}
The RT scheme further improves the PR scheme by introducing retransmission. Recall that there is no feedback provided by the destination, the source thus keeps retransmitting the current status update until a new status update is generated at the source. When a new status update is generated at the source, the source starts the service of the new status update. During the retransmission, the destination discards repetitive status updates if it has already decoded the status update correctly.
\subsection{The Evolution of AoI}
\begin{figure}[t]
\centering {\scalebox{0.35}{\includegraphics{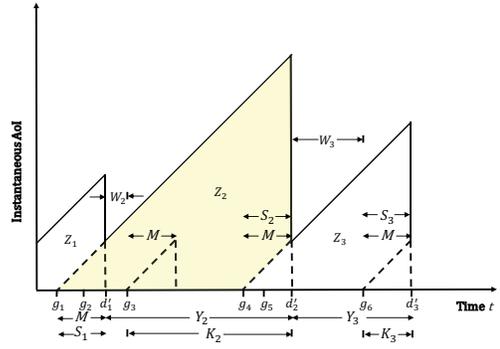}}}
\caption{The evolution of the AoI for the non-preemption scheme. }\label{fig:queueing1}
\end{figure}

In this subsection, we first give the definition of the instantaneous AoI and then define some important intervals in order to derive the average AoI. Note that the following definitions can be used for all the three proposed schemes. As an example, we depict the evolution of AoI for the NP scheme in Fig. \ref{fig:queueing1}. Due to the limited space, we will not provide the figures for PR and RT schemes.
We let $r\left(t\right)$ be the most recently received status update at time $t$, the instantaneous AoI of the considered NP, PR and RT schemes can be expressed as
\begin{equation}\label{IAoI}
\mathord{\buildrel{\lower3pt\hbox{$\scriptscriptstyle\frown$}}
\over \Delta} = t-r\left(t\right).
\end{equation}
We let $g_j, j=1,2,3,\cdots$, denotes the generation time of the $j$-th update. We define $X_j={g_{j+1}}-{g_{j}}$ as the interarrival time between consecutive updates. Since the status updates are generated according to a Poisson process, $X_j$ follows an exponential distribution with the probability density function (PDF) given by ${f_{X_j}}\left( x \right)=\lambda \exp \left({ - \lambda x} \right)$. Note that the generated status update may not be received correctly by the destination due to transmission error, discard and preemption. We let $d'_i$ be the departure time and $S_i$ be the service time of the $i$-th status update that is successfully decoded at the destination, e.g., the generated status update at $g_5$ is discarded and the correctly received status update at $d'_2$ is the status update generated at $g_4$ in Fig. \ref{fig:queueing1}.

We denote $W_i$ as the waiting time from the reception of the $\left(i-1\right)$th status update until the generation of the next status update and it is defined as
\begin{equation}
W_i={{G_i}-{d'_{i-1}}},
\end{equation}
where $G_i$ is the generation time of the first generated status update after $d'_{i-1}$ and it is defined as
\begin{equation}\label{alphai}
{G_i} \buildrel \Delta \over = \min \left\{ {{g_j}\left| {{g_j} > {d'_{i-1}}} \right.} \right\}.
\end{equation}
In addition, we define $K_i$ to be the interval starts from $G_i$ until the status update is successfully received by the destination and it is defined as
\begin{equation}
K_i={d'_i}-{G_i}.
\end{equation}
$Y_i$ is defined as the interdeparture time between two consecutive successfully received status updates and it is given by $Y_i={d'_i}-{d'_{i-1}}$. From the definitions of $W_i$ and $K_i$, $Y_i$ can be alternatively expressed as $Y_i = {W_i}+{K_i}$.

In order to derive the average AoI for the NP, PR and RT schemes, we let $N_t=\max \{ i\left| {d'_i \le t} \right.\} $ to be the number of successfully received status updates during the time interval $\left[0,t\right]$. The average AoI of the considered system can be characterized as the sum of the geometric areas $Z_i$ under the instantaneous age curve, which is given by
\begin{equation}\label{averageaoi}
\Delta  = \mathop {\lim }\limits_{t  \to \infty } {{{N_t }} \over t }{1 \over {{N_t }}}\sum\limits_{i = 1}^{{N_t }} {{Z_i}}  = \mathop {\lim }\limits_{t  \to \infty } {{{N_t }} \over t }\mathbb {E} \left\{ {{Z_i}} \right\}.
\end{equation}
With the definition of $Y_i$ and $S_i$, the area of $Z_{i}$ can be calculated as ${Z_{i}} = {{{{\left( {{Y_{i}} + {S_{i-1}}} \right)}^2}} \over 2} - {{{S_{i }}^2} \over 2}$. Since $Y_i$ and $S_{i-1}$ are independent to each other and the sequences $\left\{Y_1, Y_2,\cdots \right\}$ and $\left\{S_1, S_2,\cdots \right\}$ form independent and identically distributed processes \cite{AOIerror, MG11HARQ}, we now drop the subscript index of the intervals and obtain $\mathbb {E} \left\{ {{Z}} \right\}={{\mathbb {E}\left\{ {{Y^2}} \right\}} \over 2}+{\mathbb {E}\left\{ {{Y}} \right\}}{\mathbb {E}\left\{ {{S}} \right\}}$. Together with the fact that $ \mathop {\lim }\limits_{t  \to \infty } {{{N_t }} \over t }= {1 \over {\mathbb {E}\left\{ Y \right\}}}$, (\ref{averageaoi}) can be finally simplified to
\begin{equation}\label{delta}
\Delta  = {{\mathbb {E}\left\{ {{Y^2}} \right\}} \over {2 \mathbb {E} \left\{ Y \right\}}} + {\mathbb {E}\left\{ {{S}} \right\}}.
\end{equation}
Note that the analysis provided in this paper is readily extended to the average peak AoI, which is given by $\Delta_{peak}=\mathbb {E} \left\{ Y \right\} +\mathbb {E} \left\{ S \right\}$.

\section{Performance Analysis of Average AoI}
In this section, we evaluate the average AoI of the proposed NP, PR and RT schemes by deriving the three terms ${\mathbb {E}\left\{ {{S}} \right\}}$, ${\mathbb {E}\left\{ {{Y}} \right\}}$ and ${\mathbb {E}\left\{ {{Y^2}} \right\}}$ in (\ref{delta}) for them. Based on the derived expressions, we then minimize the average AoI for the three schemes by obtaining the optimal packet blocklengths for them.
\subsection{Non-Preemption Scheme}
We first evaluate the term ${\mathbb {E}\left\{ {{S}} \right\}}$ for the NP scheme. Recall that each status update contains fixed $N$ bits of information and is coded into a signal with $m$ c.u. (i.e., symbols). We let $T_u$ denote the unit time for each channel use, i.e., the duration for each symbol. The service time for each status update is given by ${\mathbb {E}\left\{ {{S}} \right\}}=m T_u$. We next derive the other two terms ${\mathbb {E}\left\{ {{Y}} \right\}}$ and ${\mathbb {E}\left\{ {{Y^2}} \right\}}$ for the NP scheme. For notation simplicity, we let $M=mT_u$ hereafter.

Due to the fact that $Y=W+K$, we evaluate ${\mathbb {E}\left\{ {{Y}} \right\}}$ by deriving ${\mathbb {E}\left\{ {{W}} \right\}}$ and ${\mathbb {E}\left\{ {{K}} \right\}}$. Because each departure leaves the system empty and the memoryless of Poisson process, the interval $W$ thus has the same distribution as the interarrival interval $X$ \cite{AOIerror,MG11HARQ}. We can attain that $ \mathbb E\left\{ {{W}} \right\} ={1\over \lambda}$. We now turn to the evaluation of ${\mathbb {E}\left\{ {{K}} \right\}}$. By using a recursive method \cite{AOIerror, yifanAoI1}, ${\mathbb {E}\left\{ {{K}} \right\}}$ can be expressed as
\begin{equation}\label{Yinpreeq}
\begin{split}
\mathbb E\left\{ {{K}} \right\} &= \left( {1 - \varepsilon } \right) M + \varepsilon \mathbb E \left\{ { M+ W+\hat K} \right\},
\end{split}
\end{equation}
The first term in (\ref{Yinpreeq}) indicates the event that the first generated status update is successfully received by the destination. The second term represents the case that the first generated status update cannot be decoded correctly at the destination. If this happens, recall that $W$ and $X$ has the same distribution, the source waits for an interval $W$ for the generation of a new status update. Let $\hat K$ be the remaining process of $K$ after the generation of a new status update, which is defined as the interval from the generation of a new status update until the destination correctly receives a status update. According to the definition of $K$ given in Section II-C, we notice that the evolution of $K$ is the same as its counterpart $\hat K$, and thus ${\mathbb {E}\left\{ {{K}} \right\}}= {\mathbb {E}\left\{ {{\hat K}} \right\}}$. With this result and $ \mathbb E\left\{ {{W}} \right\} ={1\over \lambda}$, we can solve ${\mathbb {E}\left\{ {{K}} \right\}}$ from (\ref{Yinpreeq}) and it is given by $\mathbb E \left\{ {{K}} \right\} = {{{\varepsilon \over \lambda } + M} \over {1 - \varepsilon }}$. $\mathbb E\left\{ {{Y}} \right\}$ can then be obtained as
\begin{equation}\label{Yinpre}
\mathbb E\left\{ {{Y}} \right\} = {\mathbb {E}\left\{ {{W}} \right\}}+{\mathbb {E}\left\{ {{K}} \right\}}={{{1 \over \lambda } + M} \over {1 - \varepsilon }}.
\end{equation}

We next derive the expectation of $Y^2$. Due to the independence between $W$ and $K$, we can obtain that
\begin{equation}\label{EY2}
\mathbb E\left\{ {{Y^2}} \right\}=\mathbb E\left\{ {{W^2}}\right\}+2\mathbb E\left\{ {{W}} \right\}\mathbb E\left\{ {{K}} \right\}+\mathbb E\left\{ {{K^2}} \right\}.
\end{equation}
Since $W$ follows an exponential distribution with rate $\lambda$, we have $\mathbb E\left\{ {W}^2 \right\} = {2 \over \lambda^2}$. Besides, the terms $E\left\{ {{W}} \right\}$ and $E\left\{ {{K}} \right\}$ have been derived above, in order to evaluate $\mathbb E\left\{ {{Y^2}} \right\}$, we need to obtain the expectation of ${K^2}$. By using a similar recursive method as in (\ref{Yinpreeq}), we have
\begin{equation}\label{NPEK2}
\begin{split}
\mathbb E\left\{ {{K^2}} \right\} &= \left( {1 - \varepsilon } \right) M^2 + \varepsilon \mathbb E \left\{ { M^2+ W^2+ K^2+}\right. \\
&\quad \left.{2MW+2MK+2WK} \right\}.
\end{split}
\end{equation}
Substituting the derived results of $\mathbb E\left\{ {{W}} \right\}$, $\mathbb E\left\{ {{K}} \right\}$ and $\mathbb E\left\{ {{W^2}} \right\}$ into (\ref{NPEK2}), we can obtain that
\begin{equation}
\mathbb E\left\{ {{K}^2} \right\} = \frac{{{\lambda ^2}{M^2} + {\lambda ^2}{M^2}\varepsilon  + 4\lambda M\varepsilon  + 2\varepsilon }}{{{\lambda ^2}{{(1 - \varepsilon )}^2}}}.\\
\end{equation}
With the derived expectations of $W$, $K$, ${W}^2$ and $K^2$, according to (\ref{EY2}), the term $\mathbb E\left\{ {{Y^2}} \right\}$ can be evaluated as
\begin{equation}\label{Yinpre2}
\mathbb E\left\{ {{Y^2}} \right\}={{{{\left( {M + {1 \over \lambda }} \right)}^2}\left( {1 + \varepsilon } \right)} \over {{{\left( {1 - \varepsilon } \right)}^2}}} + {1 \over {{\lambda ^2}\left( {1 - \varepsilon } \right)}}.
\end{equation}
Substituting ${\mathbb {E}\left\{ {{S}} \right\}}=M$, (\ref{Yinpre}) and (\ref{Yinpre2}) into (\ref{delta}), the average AoI for the NP scheme can be finally expressed as
\begin{equation}\label{AOInpre}
\Delta_{NP}  = {{\left( {M + {1 \over \lambda }} \right)\left( {1 + \varepsilon } \right)} \over {2\left( {1 - \varepsilon } \right)}} + {1 \over {2{\lambda ^2}\left( {M + {1 \over \lambda }} \right)}}+M.
\end{equation}
\subsection{Preemption Scheme}
In the PR scheme, we first note that the service time of the PR scheme is the same as NP scheme since retransmission is not considered, i.e., ${\mathbb {E}\left\{ {{S}} \right\}}=M$. We next evaluate the term $\mathbb E\left\{ {{Y}} \right\}= \mathbb E\left\{ {{W} } \right\} +\mathbb E\left\{ { {K}} \right\}$ by deriving $\mathbb E\left\{ {{W} } \right\} $ and $\mathbb E\left\{ {{K} } \right\} $. We also have $\mathbb E\left\{ {{W} } \right\}={1\over \lambda}$ and the expectation of $K$ for the PR scheme can be derived as
\begin{equation}\label{Kieq}
\begin{split}
\mathbb E\left\{ { {K}} \right\}& =  \underbrace {\left(1-p\right)\left(1-\varepsilon\right)M}_{L_1}\\
& \quad +\underbrace {\left(1-p\right)\varepsilon\left(M+\mathbb E\left\{ { {W}} \right\}+ \mathbb E\left\{ { K} \right\}\right)}_{L_2}\\
&\quad +\underbrace {p\left(\mathbb E\left\{ {X\left| {X < M} \right.} \right\}+\mathbb E\left\{ { {K}} \right\}\right)}_{L_3}.
\end{split}
\end{equation}
In the above expression, the term $L_1$ denotes the case that the first generated status update is not preempted by other status updates and is decoded successfully by the destination. The length of the interval $K$ is just the service time $M$. The term $1-p$ denotes the probability that the current status update is not preempted, and the preemption probability $p$ can be evaluated as
\begin{equation}\label{p}
p = \Pr \left\{ {{X} < M} \right\} = 1 - \exp \left( { - \lambda M} \right).
\end{equation}
The term $L_2$ indicates the case that the first generated status update is not preempted and cannot be decoded correctly by the destination. For such case, the system first spends an interval $M$ on the first status update, then waits for a period $W$ until the generation of a next update. After the generation of a new status update, the remaining process is the same as $K$, which is illustrated after (\ref{Yinpreeq}). The term $L_3$ shows the case that the first generated status update is preempted by a new status update. The conditional expectation $\mathbb E\left\{ {X\left| {X < M} \right.} \right\}$ is the generation interval under the condition that the status update is preempted. The system then follows the same process as $K$ after preemption. The conditional expectation in (\ref{Kieq}) can be derived as
\begin{equation}\label{expect}
\begin{split}
\mathbb E\left\{ {X\left| {X < M} \right.} \right\} &= {{\int_0^M {x\lambda {e^{ - \lambda x}}dx} } \over {1 - {e^{ - \lambda M}}}}= {1 \over \lambda } + M - {M \over p},
\end{split}
\end{equation}
where the integral is solved by \cite[Eq. 3.351-1.8]{Tableofintegral}. Substituting $\mathbb E\left\{ {{W}} \right\}={1\over \lambda}$, (\ref{p}) and (\ref{expect}) into (\ref{Kieq}), we can solve that
\begin{equation}\label{Ki}
\mathbb E\left\{ { {K}} \right\}= {{{p + \varepsilon  - p\varepsilon }} \over {\lambda \left( {1 - p} \right)\left( {1 - \varepsilon } \right)}}.
\end{equation}
The expectation of interval $Y$ for the PR scheme can thus be evaluated as
\begin{equation}\label{Yipre}
\begin{split}
\mathbb E\left\{ {{Y}} \right\} & =  \mathbb E\left\{ { {W}} \right\}+ \mathbb E\left\{ { {K}} \right\}={1 \over {\lambda \left( {1 - p} \right)\left( {1 - \varepsilon } \right)}} .\\
\end{split}
\end{equation}
To calculate the expectation of $Y^2$ for the PR scheme, we follow (\ref{EY2}) and evaluate the expectation of $K^2$. Note that the other terms in (\ref{EY2}) for the PR scheme have been derived in this subsection. The equation for $\mathbb E\left\{ { {K^2}} \right\}$ of the PR scheme is characterized in (\ref{Ki2eq}) given on top of the next page.
\begin{figure*}[!t]
\begin{equation}\label{Ki2eq}
\begin{split}
\mathbb E\left\{ { {K^2}} \right\}& =  \left(1-p\right)\left(1-\varepsilon\right)M^2+\left(1-p\right)\varepsilon \mathbb E \left\{ { M^2+ W^2+K^2+2MW+2MK+2WK} \right\}+\\
&\quad p\left(\mathbb E\left\{ {X^2\left| {X < M} \right.} \right\}+2\mathbb E\left\{ {X\left| {X < M} \right.} \right\}\mathbb E\left\{ { {K}} \right\}+\mathbb E\left\{ { {K^2}} \right\}\right)
\end{split}
\end{equation}
\hrulefill
\end{figure*}
With the help of \cite[Eq. 3.351-1.8]{Tableofintegral}, we have
\begin{equation}\label{expect1}
\begin{split}
\mathbb E\left\{ {X^2\left| {X < M} \right.} \right\} &= {{\int_0^M {x^2\lambda {e^{ - \lambda x}}dx} } \over {1 - {e^{ - \lambda M}}}}\\
& = {2 \over {{\lambda ^2}}} + {{2M} \over \lambda } + {M^2} - {{{{2M} \over \lambda } + {M^2}} \over p}.
\end{split}
\end{equation}
Substituting $\mathbb E\left\{ { {W}} \right\} = {1\over \lambda}$, $\mathbb E\left\{ { {W^2}} \right\}={2\over \lambda^2}$, (\ref{expect}), (\ref{Ki}) and (\ref{expect1}) into (\ref{Ki2eq}), we can derive that
\begin{equation}\label{Ki2eqd}
\begin{split}
\mathbb E\left\{ { {K^2}} \right\} &=  {{2\left( {p + \varepsilon  - p\varepsilon } \right)} \over {\lambda \left( {1 - p} \right)^2\left( {1 - \varepsilon } \right)^2}}\left( {{1 \over \lambda } + M} \right)\\
&\quad- {{2M} \over {\lambda \left( {1 - p} \right)^2\left( {1 - \varepsilon } \right)^2}}.\\
\end{split}
\end{equation}
With the derived expectations, according to (\ref{EY2}), $\mathbb E\left\{ {{Y^2}} \right\}$ can be evaluated as
\begin{equation}\label{Yi2pre}
\begin{split}
\mathbb E\left\{ {{Y^2}} \right\}={2 \over {{\lambda ^2}{{\left( {1 - p} \right)}^2}{{\left( {1 - \varepsilon } \right)}^2}}} - {{2M} \over {\lambda \left( {1 - p} \right)\left( {1 - \varepsilon } \right)}}.
\end{split}
\end{equation}
Substituting ${\mathbb {E}\left\{ {{S}} \right\}}=M$, (\ref{Yipre}), and (\ref{Yi2pre}) into (\ref{delta}), we obtain the average AoI for the PR scheme given by
\begin{equation}\label{PR}
\Delta_{PR}  = {1 \over {\lambda \exp \left( { - \lambda M} \right)\left( {1 - \varepsilon } \right)}}.
\end{equation}
\subsection{Retransmission Scheme}
We now study the average AoI of the RT scheme. We first characterize the expectation of $S$ for the RT scheme. Different from the NP and PR schemes that the service time is a fixed value, e.g., $S=M$, the service time of the RT scheme is a multiple of $M$ because the status update may be decoded correctly after some rounds of retransmissions. Specifically, the expectation of $S$ can be characterized as ${\mathbb {E}\left\{ {{S}} \right\}}={E_S \over p_S}$, where $E_S$ is the expectation of service time for the event that the considered status update is not preempted by new updates and finally decoded successfully by the destination and $p_S$ denotes the total probability of such event. $E_S$ and $p_S$ can be evaluated as
\begin{equation}\label{ESeq}
{E_S} = {\sum\limits_{k = 0}^\infty  {{\varepsilon ^k}\left( {1 - \varepsilon } \right)\left( {1 - p} \right)} ^{k + 1}}\left( {k + 1} \right)M,
\end{equation}
\begin{equation}\label{PSeq}
{p_S} = {\sum\limits_{k = 0}^\infty  {{\varepsilon ^k}\left( {1 - \varepsilon } \right)\left( {1 - p} \right)} ^{k + 1}},
\end{equation}
where $p$ given in (\ref{p}) is the probability that an update being served is preempted by a new update and $k$ denotes the number of retransmission times. Applying the series expansion ${1 \over {1 - x}} = \sum\limits_{n = 0}^\infty  {{x^n}} $ and ${1 \over {{{\left( {1 - x} \right)}^2}}} = \sum\limits_{n = 0}^\infty  {\left( {n + 1} \right){x^n}} $, we can simplify the expression of $E_S$ and $p_S$ to $E_S={{\left( {1 - p} \right)\left( {1 - \varepsilon } \right)M} \over {{{\left[ {1 - \varepsilon \left( {1 - p} \right)} \right]}^2}}}$ and $p_S={{\left( {1 - p} \right)\left( {1 - \varepsilon } \right)} \over {1 - \varepsilon \left( {1 - p} \right)}}$. We can now derive the expectation of the service time for the RT scheme given by
\begin{equation}\label{Sire}
{\mathbb {E}\left\{ {{S}} \right\}}={M \over {1 - \varepsilon \left( {1 - p} \right)}}.
\end{equation}

The other two expectation terms $\mathbb E\left\{ { {Y}} \right\}$ and $\mathbb E\left\{ { {Y^2}} \right\}$ for the RT scheme can be derived similarly as those for NP and PR schemes analyzed in Section III-A, B. Due to the limited space, we do not provide the detailed derivation of $\mathbb E\left\{ { {Y}} \right\}$ and $\mathbb E\left\{ { {Y^2}} \right\}$. The final expressions of $\mathbb E\left\{ { {Y}} \right\}$ and $\mathbb E\left\{ { {Y^2}} \right\}$ for the RT scheme are given by
\begin{equation}\label{Yire}
\mathbb E\left\{ {{Y}} \right\} ={1-\varepsilon+p\varepsilon  \over {\lambda \left( {1 - p} \right)\left( {1 - \varepsilon } \right)}},
\end{equation}
\begin{equation}\label{Yire2}
\begin{split}
\mathbb E\left\{ { {Y^2}} \right\} &=2{\left( {{{1 - \varepsilon  + p\varepsilon } \over {\lambda \left( {1 - p} \right)\left( {1 - \varepsilon } \right)}}} \right)^2} - {{2M} \over {\lambda \left( {1 - p} \right)\left( {1 - \varepsilon } \right)}}.
\end{split}
\end{equation}
Substituting (\ref{Sire}), (\ref{Yire}) and (\ref{Yire2}) into (\ref{delta}), we can derive the average AoI for the retransmission scheme given by
\begin{equation}\label{RT}
\Delta_{RT}={{1 - \varepsilon \exp \left( { - \lambda M} \right)} \over {\lambda \exp \left( { - \lambda M} \right)\left( {1 - \varepsilon } \right)}}.
\end{equation}
\subsection{Optimal Packet Blocklength}
We realize that there exists optimal packet blocklengths for the three schemes such that their average AoI is minimized. Specifically, a smaller system blocklength may potentially decrease the average AoI for the NP, PR and RT schemes because the unit $M=mT_u$ is reduced. However, this may also increase the average AoI because the BLER $\varepsilon$ is increased by reducing $m$. Due to the complicated structure of $\varepsilon$ in (\ref{BLER}), it is hard to characterize closed-form expressions of the optimal packet blocklengths for the three proposed schemes. Alternatively, we derive three equations of the optimal values of blocklengths for the three schemes, where the optimal $m$ can be easily solved by numerical methods.

For the NP scheme, the optimal packet blocklength is the solution to the equation ${{d\Delta_{NP} } \over {dm}}=0$, which is given by
\begin{equation}\label{opmnpre}
\begin{split}
{T_u}\left( {{{1 - {\varepsilon ^2}} \over {2{{\left( {1 - \varepsilon } \right)}^2}}} + 1} \right) + {{\left( {{T_u}m + {1 \over \lambda }} \right)} \over {{{\left( {1 - \varepsilon } \right)}^2}}}{{\partial \varepsilon } \over {\partial m}}= {{{T_u}} \over {2{\lambda ^2}{{\left( {{T_u}m + {1 \over \lambda }} \right)}^2}}}.\\
\end{split}
\end{equation}
According to (\ref{BLER}) and the fact that ${{\partial Q\left( x \right)} \over {\partial x}} =  - {{\exp \left( { - {{{x^2}} \over 2}} \right)} \over {\sqrt {2\pi } }}$, the term ${{\partial \varepsilon } \over {\partial m}}$ in (\ref{opmnpre}) is given by
\begin{equation}\label{dedm}
\begin{split}
{{\partial \varepsilon } \over {\partial m}}=-{{\exp \left( -{{{{\Psi ^2}} \over 2}} \right)} \over {\sqrt {2\pi } }}{{\sqrt 2 \left( {{N \over {m\sqrt m }} + {{{{\log }_2}\left( {1 + \gamma } \right)} \over {2\sqrt m }}} \right)} \over {2{{\log }_2}\left( e \right)\sqrt {1 - {1 \over {{{\left( {1 + \gamma } \right)}^2}}}} }},
\end{split}
\end{equation}
where $\Psi={{{{1 \over 2}{{\log }_2}\left( {1 + \gamma } \right) - {N \over m}} \over {{{\log }_2}\left( e \right)\sqrt {{1 \over {2m}}\left( {1 - {1 \over {{{\left( {1 + \gamma } \right)}^2}}}} \right)} }}}$. The optimal blocklength of the NP scheme can be obtained by solving $m$ from (\ref{opmnpre}). Next, for the PR scheme, the optimal blocklength is the solution to the equation ${{d\Delta_{NP} } \over {dm}}=0$, which is given by
\begin{equation}
\lambda {T_u}\left( {1 - \varepsilon } \right) + {{\partial \varepsilon } \over {\partial m}} = 0.
\end{equation}
Based on (\ref{RT}), the equation used for the RT scheme to solve the optimal blocklength is given by
\begin{equation}
\lambda {T_u}\left( {1 - \varepsilon } \right) + {{\partial \varepsilon } \over {\partial m}}\left( {1 - {e^{ - \lambda {T_u}m}}} \right) = 0.
\end{equation}

\section{Numerical Results}
In this section, we present some numerical results to validate and illustrate the above theoretical analysis. Specifically, we set the information of each status update $N=150$ bits and the unit time for each channel use (i.e., symbol duration) $T_u=0.006$. The setting of the unit time is chosen to map the service time in unit time, i.e., $M$, to the generation interval $X$ such that their values are in the same order of magnitude. For example, for a system with $m=200$ and $\lambda=0.33$, the service time for each transmission is 1.2 unit time and the average generation interval is 3 unit time.
\begin{figure}
\centering
  {\scalebox{0.5}{\includegraphics {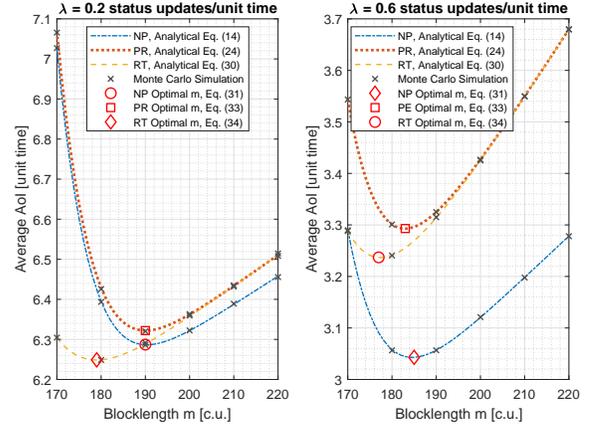}
  \label{fig:mplot1}}}
\caption{The average AoI versus packet blocklength $m$ for different generation rates, where the SNR $\gamma=4.5$ dB.
\label{montsimu}}
\end{figure}

\begin{figure}
\centering
  {\scalebox{0.5}{\includegraphics {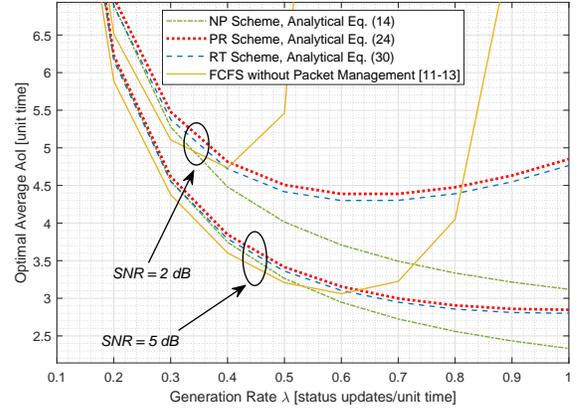}
  \label{fig:mplot2}}}
\caption{The optimal average AoI versus the generation rate $\lambda$ with optimal packet blocklength.
\label{montsimu1}}
\end{figure}
We first depict the average AoI versus the packet blocklength for NP, PR and RT schemes in Fig. \ref{montsimu}. We can observe that all the analytical results coincide well with the simulations, which validates the theoretical analysis provided in Section III. Besides, the RT scheme outperforms the PR scheme in both considered setups. The rationale behind the performance gain is that the RT scheme uses the idle slots to retransmit the packets, which increases the reliability and reduces the average AoI. Moreover, the NP scheme and the RT scheme can outperform each other depending on different generation rate $\lambda$. Specifically, the NP scheme is preferable for a system with a relatively large generation rate and the RT scheme is suitable for a system with a small value of $\lambda$. This observation can be explained as follows. In the NP scheme, the average AoI decreases as $\lambda$ increases because the waiting time after a successful transmission is reduced. Specifically, when $\lambda \to 1$, the source immediately transmits another status update to the destination after finishing the service of last status update. Differently, in the RT scheme, a smaller value of $\lambda$ decreases the probability that a packet is being preempted. By keeping retransmitting the status updates instead of keeping idle, RT scheme can outperform the NP scheme when $\lambda$ is low. However, when $\lambda$ is high, the service of each status updates may be preempted frequently before it is finished, which further decreases the average AoI. As the analytical results agree with the simulation results well, we only plot the analytical results of the NP, PR and RT schemes in the next figure.

Fig. \ref{montsimu1} illustrates the optimal average AoI versus the generation rate with the optimal settings of packet blocklength. In Fig. \ref{montsimu1}, the optimal values of packet blocklength for the three proposed schemes are set to the analytical results derived in (31), (33) and (34). The FCFS without packet management is the scheme considered in \cite{AoISP,AoISP1,AoISP2}. Specifically, the buffer of the FCFS scheme without packet management has an infinite size, and the source keeps retransmitting each status updates until it is successfully received by the destination in an FCFS order. The packet blocklength of the FCFS without packet management is optimized by a one-dimensional exhaustive search. We can observe from Fig. \ref{montsimu1} that the proposed three schemes outperform the existing FCFS without packet management scheme for a large range of generation rate. This is thanks to the introduction of the packet management strategies in terms of preemption and discard. Without packet management, the scheme considered in \cite{AoISP,AoISP1,AoISP2} becomes unstable when the generation rate is high because the queue in the buffer grows unbounded. Moreover, the scheme considered in \cite{AoISP,AoISP1,AoISP2} slightly outperforms the proposed schemes at the low generation rate regime. This is because in our considered model, the buffer can only store one status update. There exists an additional waiting period elapsed since the successful transmission of a status update until the system generates a new status update again.


%
\section{Conclusions}
In this paper, we considered a point-to-point short-packet communication system where the source randomly generates status updates and transmits the status updates to the destination as timely as possible. The status updates are assumed to contain packets with limited size and thus the finite blocklength coding is implemented. Based on the considered model, we proposed three packet management schemes, namely NP, PR and RT schemes, to timely deliver the generated status updates. To characterize the timeliness of the status updates, we adopt a new performance metric named AoI and derived the average AoI for the proposed schemes. Based on the derived expressions, we then optimized the optimal packet blocklength of the proposed schemes in terms of minimizing their corresponding average AoI. Simulation results validated all the theoretical analysis and showed that the proposed schemes can outperform each other for different system setups. More importantly, the proposed schemes outperform the existing scheme FCFS without packet management when the generation rate of the source is not very low.
\appendices
\ifCLASSOPTIONcaptionsoff
  \newpage
\fi

\bibliographystyle{IEEEtran}
\bibliography{References}

\end{document}